\documentclass[prl,twocolumn,english,superscriptaddress,floatfix,longbibliography]{revtex4}

\usepackage{graphicx}
\usepackage{bm, amsmath, amssymb}
\usepackage{xcolor}
 \usepackage{soul}
\usepackage{pdfpages}
\usepackage[T1]{fontenc}
 \usepackage{babel}

\usepackage{amstext}
\usepackage{bbold}
\usepackage{esint}

\usepackage{babel}

\begin{document}

\title{Information gain and loss for a quantum Maxwell's demon}

\author{M. Naghiloo}
\affiliation{Department of Physics, Washington University, St.\ Louis, Missouri 63130}
\author{J. J. Alonso}
\affiliation{Department of Physics, Friedrich-Alexander-Universit\"at Erlangen-N\"urnberg, D-91058 Erlangen, Germany}
\author{A. Romito}
\affiliation{Department of Physics, Lancaster University, Lancaster LA1 4YB, United Kingdom}
\author{E. Lutz}
\affiliation{Department of Physics, Friedrich-Alexander-Universit\"at Erlangen-N\"urnberg, D-91058 Erlangen, Germany}
\author{K. W. Murch}
\affiliation{Department of Physics, Washington University, St.\ Louis, Missouri 63130}
\affiliation{Institute for Materials Science and Engineering, St.\ Louis, Missouri 63130}

\begin{abstract}
We use continuous weak measurements of a driven superconducting qubit to experimentally study the information dynamics of a quantum Maxwell's demon. We show how information gained by a demon who can track single quantum trajectories of the qubit  can be converted into work using quantum coherent feedback. We verify the validity of a quantum fluctuation theorem with feedback by utilizing information obtained along single trajectories.  We demonstrate, in particular, that quantum backaction  can lead to a loss of information in imperfect measurements. We furthermore probe the transition between information gain and loss by varying the initial purity of the qubit. 
\end{abstract}
\maketitle
\vspace{.5pc}


The thought experiment of Maxwell's demon reveals the profound connection between information and energy in thermodynamics \cite{bril56,maru09,leff14,parr15,lut15}. By knowing the positions and velocities of each molecule in a gas, the demon can sort hot and cold particles without performing any work, in apparent violation of the second law. Thermodynamics must therefore be generalized to incorporate information in a consistent manner. Classical Maxwell's demon experiments have been realized with cold atoms \cite{raiz09}, a molecular ratchet \cite{serr07}, colloidal particles \cite{toya10,rold14}, single electrons \cite{kosk14a,kosk14} and photons \cite{vidr16}. Recent advances in fabrication and control of small systems where quantum fluctuations are dominant over thermal fluctuations allow for novel studies of quantum thermodynamics \citep{bata14,bata15,an15,nagh17,ceri17,smit18,xion18}. In particular, Maxwell's demon has been realized in several systems using feedback control to study the role of information in the quantum regime \cite{cama16,ciam17,cott17,masu17}. While these experiments probe information and energy dynamics in the regime of single energy quanta, the dynamics either does not include quantum coherence  or the demon destroys these coherences through projective measurements \cite{cama16,ciam17,cott17,masu17}. Therefore, in either case, the action of the demon can be understood using entirely classical information. However, in quantum systems the information exchanged in a measurement may present strikingly nonclassical features owing to the measurement backaction \cite{groe71,jaco06a,jaco14,funo13}.

In this work, we use continuous weak measurements followed by feedback control of a superconducting qubit \cite{wall04,hatr13,murc13traj,webe14} to realize Maxwell's demon in a truly quantum situation, where  quantum backaction and quantum coherence contribute to the dynamics. This approach enables us  to experimentally verify a quantum fluctuation theorem with feedback \cite{saga08,mori11,funo13} at  the level of single quantum trajectories. This fluctuation theorem is a  nonequilibrium extension of the second law that accounts  for both quantum fluctuations and the information collected by the demon. At the same time, this method allows us to study the role of quantum backaction and quantum coherence in the acquired information. In particular, we show that the average  information exchanged with the detector can be negative due to measurement backaction. Here  the loss of information associated with the perturbing effect of the detector  dominates the measurement process.  By preparing the qubit at different temperatures according to a Gibbs distribution, we experimentally map out the full transition between regimes of information gain  and information loss \cite{funo13}. 

\begin{figure}
\centering
\includegraphics[scale=0.9]{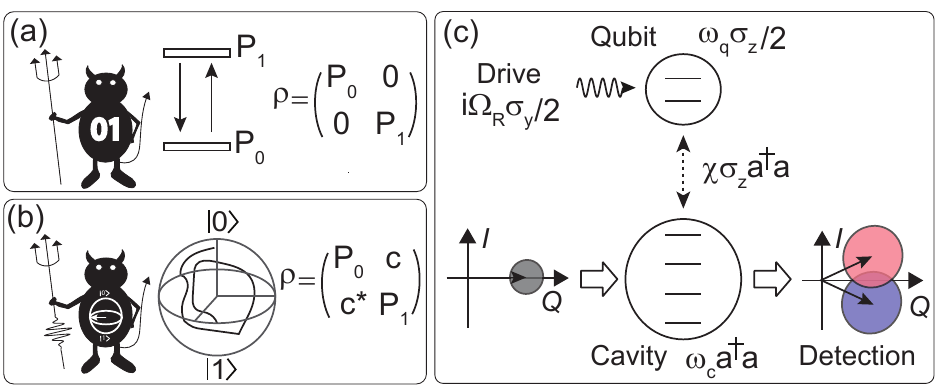}
\caption{Classical demon vs quantum demon. (a) In the classical situation the dynamics can be seen  as an evolution of the populations in the definite eigenstates, yet in the quantum case (b), the dynamics includes coherences and can no longer  be understood as a classical mixture. (c)  The experimental configuration consists of a quantum two-level system 
coupled to a cavity mode 
 via a dispersive interaction
   which allows for both weak and strong measurements of the qubit state populations. A resonant drive at the qubit transition frequency 
   turns these populations into coherences and vice versa leading to coherent quantum evolution.} 
\label{fig1}
\end{figure}

\textit{Superconducting circuit}---In order to study the information exchanged with the detector in a genuinely  quantum situation, we employ quantum measurement techniques in a superconducting circuit to realize a quantum Maxwell's demon (Fig.~1a,b). Our setup consists of a  transmon qubit dispersively coupled to a 3D aluminum cavity
by which we readout the state of the qubit with dispersive measurement using Josephson parametric amplifier operating in phase sensitive mode with a total measurement quantum efficiency of 30\% (Fig.~\ref{fig1}c). 
The corresponding effective Hamiltonian in the presence of a coherent drive is,
\begin{eqnarray} \label{Ham}
H & =& -\frac{\omega_\mathrm{q}}{2} \sigma_z  -i \Omega_\mathrm{R} \sigma_y \cos(\omega_\mathrm{q} t) - \chi a^{\dagger}a\sigma_z+ \omega_\mathrm{c} a^{\dagger}a,\ \ 
\end{eqnarray}
where $\sigma_z, \sigma_y$ are Pauli matrices, $\omega_\mathrm{q}$  the resonance frequency of the qubit, $\omega_\mathrm{c}$ is the cavity frequency, $a^\dagger, a$ are creation and annihilation operators, and $\Omega_\mathrm{R}$ is the Rabi drive frequency \cite{sampleparameters}. The quantity $\chi$ is the dispersive coupling rate between the cavity mode and the qubit state. In this measurement architecture \cite{wall04,hatr13,murc13traj,webe14}, the qubit-state-dependent phase shift of a weak cavity probe tone is continuously monitored, resulting in a measurement record, $r$, that is proportional to $\langle\sigma_z\rangle$.  Using the record $r$ we obtain the qubit conditional state evolution $\rho_{t|r}$, which depends on the record from time $0$ to $t$, using the stochastic master equation (SME) \cite{tan15,foro16},
\begin{eqnarray} \label{SME}
\dot{\rho}_{t|r} & =& \frac{1}{i\hbar}[H_R,\rho_{t|r}] + k (\sigma_z \rho_{t|r} \sigma_z  - \rho_{t|r})\nonumber \\
 &+& 2 \eta k (\sigma_z \rho_{t|r} + \rho_{t|r} \sigma_z - 2 \mathrm{Tr}(\sigma_z \rho_{t|r})\rho_{t|r}) r(t),
\end{eqnarray}
where $k$ is the strength of the measurement, $\eta$ is the efficiency of the detector, and $H_\mathrm{R} = -i \Omega_\mathrm{R}/2 \sigma_y$ describes the qubit drive in the rotating frame  \citep{sampleparameters}.  The first two terms correspond to the standard Lindblad master equation that accounts for unitary evolution and dephasing of the qubit by the dispersive measurement. The third term describes the state update due to the measurement record which includes the stochastic measurement signal $r(t) \propto \langle \sigma_z \rangle(t) + dW_t$, with $dW_t$ a zero-mean Gaussian
distributed Wiener increment \cite{jaco06}. Owing to the weak coupling to the measuring device, information about the state of the qubit may be  gathered without projecting it into energy eigenstates, thus preserving coherent superpositions.

\textit{The demon's information}---The SME \eqref{SME}  tracks the state of knowledge about the qubit obtained by the quantum demon.   The amount of information exchanged with the detector  depends on both the measurement outcome and the state of the system. It may be quantified as \cite{funo13},
\begin{eqnarray} 
I(\rho_{t|r},r)&=& \ln P_{z'}(\rho_{t|r}) - \ln P_z(\rho_0)
\label{Mut_inf1},
\end{eqnarray}
where $P_{z'}$ represents the probability of getting the result $z'=0,1$ in the $z'$-basis where the system is diagonal. The stochastic evolution of the information \eqref{Mut_inf1} along a quantum trajectory follows as,
\begin{align}
\tilde{I}_r &=  \sum_{z,z'=\pm 1} [P_{z'}(\rho_{t|r}) \ln P_{z'}(\rho_{t|r}) - P_z(\rho_0)\ln P_z(\rho_0)] \nonumber \\
 &= S(\rho_0) -S(\rho_{t|r}), \label{Mut_inf2} 
\end{align}
with the von Neumann entropy $S(\rho) =-{\rm Tr} [\rho \ln \rho]$.
In Eq.~\eqref{Mut_inf2} the conditional probabilities $P_{z'}(\rho_{t|r})$ come from the SME corresponding to a \emph{single} run of the experiment, that is, an individual quantum trajectory. The averaged value of the exchanged information is obtained by averaging over many trajectories,
\begin{equation}
\langle I \rangle =\sum_r p(r) \tilde{I}_r = S(\rho_0)- \sum_r p(r) S(\rho_{t|r}),
\label{Mut_inf3}
\end{equation}
where $p(r)$ is the probability density of the measurement record $r$. 
Equation \eqref{Mut_inf3} is the information about the state of the system  gathered by the quantum demon \cite{groe71,ozaw86,jaco14}. Remarkably, it may positive or negative.
\begin{figure*}
\centering
\includegraphics[scale=0.9]{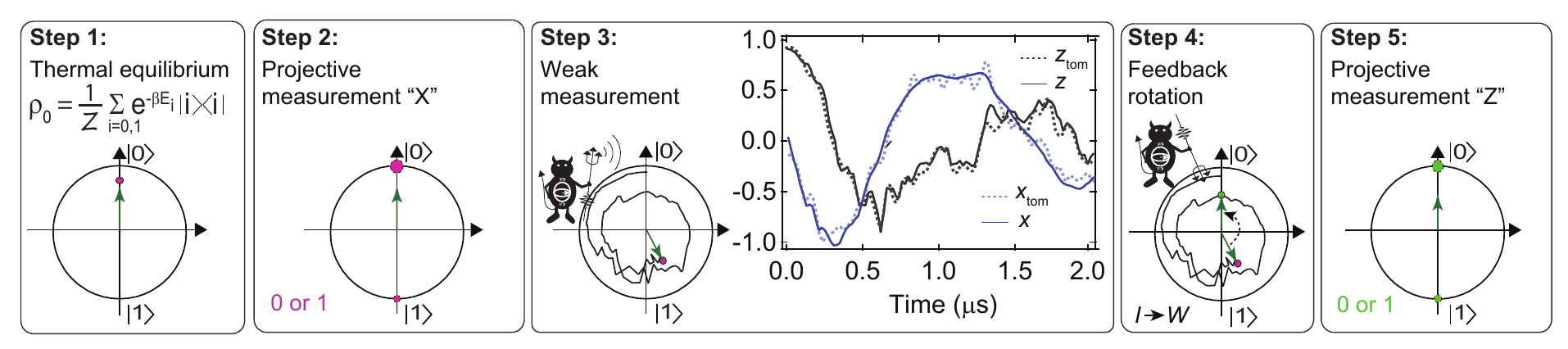}
\caption{ Experimental sequence. Step 1: The qubit is initialized in a  thermal state at inverse temperature $\beta$. Step 2: As the first step in determining the energy change  through projective measurements, we perform a  projective measurement (labeled ``X'') in the energy basis. Step 3: We drive the qubit with a coherent drive characterized by $\Omega_R/2\pi = 0.8$ MHz while the quantum demon monitors the qubit evolution with a near-quantum-limited detector. The demon's knowledge about the state can be expressed in terms of the expectation values $x\equiv \langle \sigma_x \rangle$ and $z \equiv \langle \sigma_z \rangle$ (solid lines) with the corresponding tomographic validation showing that the demon's expectation values are verified with quantum state tomography \cite{murc13traj,webe14}. Step 4: The quantum demon uses the acquired information from the previous step to apply a rotation to bring the qubit to the ground state. Step 5: We perform a projective measurement labeled ``Z'' as the second step in a two-point energy measurement. }
\label{fig2}
\end{figure*}

For classical measurements, e.g.\ when the measurement operator commutes with the state, Eq.~\eqref{Mut_inf3} reduces to the classical mutual information which is always positive \cite{saga12}.  By contrast, quantum measurements perturb the state in addition to acquiring information. Due to this unavoidable quantum backaction, the uncertainty in the detector state can be transferred to the system and increase its entropy. While Eq.~\eqref{Mut_inf3} is positive for efficient measurements, it may become negative for inefficient measurements  \cite{groe71,ozaw86,jaco14}.
Quite generally, modelling the detector uncertainty as an average over inaccessible degrees of freedom, parametrized by a stochastic variable $a$, the exchanged information \eqref{Mut_inf3} may be written as a sum of information gain and information loss, $\langle I \rangle =I_{\rm gain} - I_{\rm loss}$, with $I_{\rm gain} =S(\rho_0)- \sum_a p(a,r) S(\rho_{t|r,a}) \geqslant 0$ and  $I_{\rm loss} =\sum_r S(\rho_{t|r})- \sum_a p(a,r) S(\rho_{t|r,a}) \geqslant 0$ \cite{funo13}. Expression \eqref{Mut_inf3} is hence negative whenever the information loss induced by the quantum backaction is larger than the information acquired through the measurement.

\textit{Feedback protocol}---We now turn to the experimental procedure and our feedback protocol. The experiment consists of five steps as depicted in Figure~\ref{fig2}. In Step 1, the qubit is initialized in a given thermal state characterized by an inverse temperature $\beta$. Experimentally, we have control over $\beta$ by applying a short excitation pulse to the qubit at the start of the experimental sequence. In Step 2, we perform a first projective measurement, which, when combined with a second projective measurement during the final step, will quantify the energy change during the whole protocol \cite{talk07}. In Step 3, we employ a continuous resonant drive at the qubit transition frequency to induce Rabi oscillations of the qubit state. In conjunction with the drive, we continuously probe the cavity with a measurement rate $k/2\pi=51$~kHz generating a measurement record $r$ for the demon to track the evolution according to the SME \eqref{SME}. The axis of the resonant drive and measurement basis constrict the evolution of the qubit to the $X$--$Z$ plane of the Bloch sphere.  A typical qubit evolution is depicted by solid lines in the Step 3-inset of Fig.~\ref{fig2}. The quantum trajectory is validated with quantum state tomography as indicated by the dashed lines \cite{murc13traj,webe14}. At the final time $\tau$, the demon uses the knowledge about the state of the system to perform a rotation in Step 4 to bring the qubit back to the ground state, and extract work. To implement the feedback, we perform a random rotation pulse in the range of $[0,2\pi]$ and select the correct rotations (within the error of $\pm \pi/20$) in a post-processing step. This approach avoids long loop delays that occur for realtime feedback. We eventually finish the experiment with the second projective measurement in Step 5.  We note that the measurement basis ($\sigma_z$) is the same in Steps 2 and 5. We evaluate averaged quantities  by repeating the experiment many times.

\textit{Experimental results}---We begin by experimentally verifying  a quantum fluctuation theorem with feedback in the form of a generalized quantum Jarzynski equality, $\langle \exp{[-\beta (W- \Delta F) - I ]} \rangle =1$, where $W$ is the work done on the system by the external driving, $\Delta F$ the equilibrium free energy difference between final and initial states, and $I$ the information \eqref{Mut_inf1}. This fluctuation theorem generalizes the second law to account for quantum fluctuations and information exchange. It has been derived for classical systems in Ref.~\cite{saga08} and experimentally investigated in Refs.~\cite{toya10,kosk14}. It has later been extended to quantum systems in Refs.~\cite{mori11,funo13} and recently experimentally studied with a two-level system whose dynamics is that of a classical (incoherent) mixture \cite{masu17}.

\begin{figure}
\centering
\includegraphics[scale=0.9]{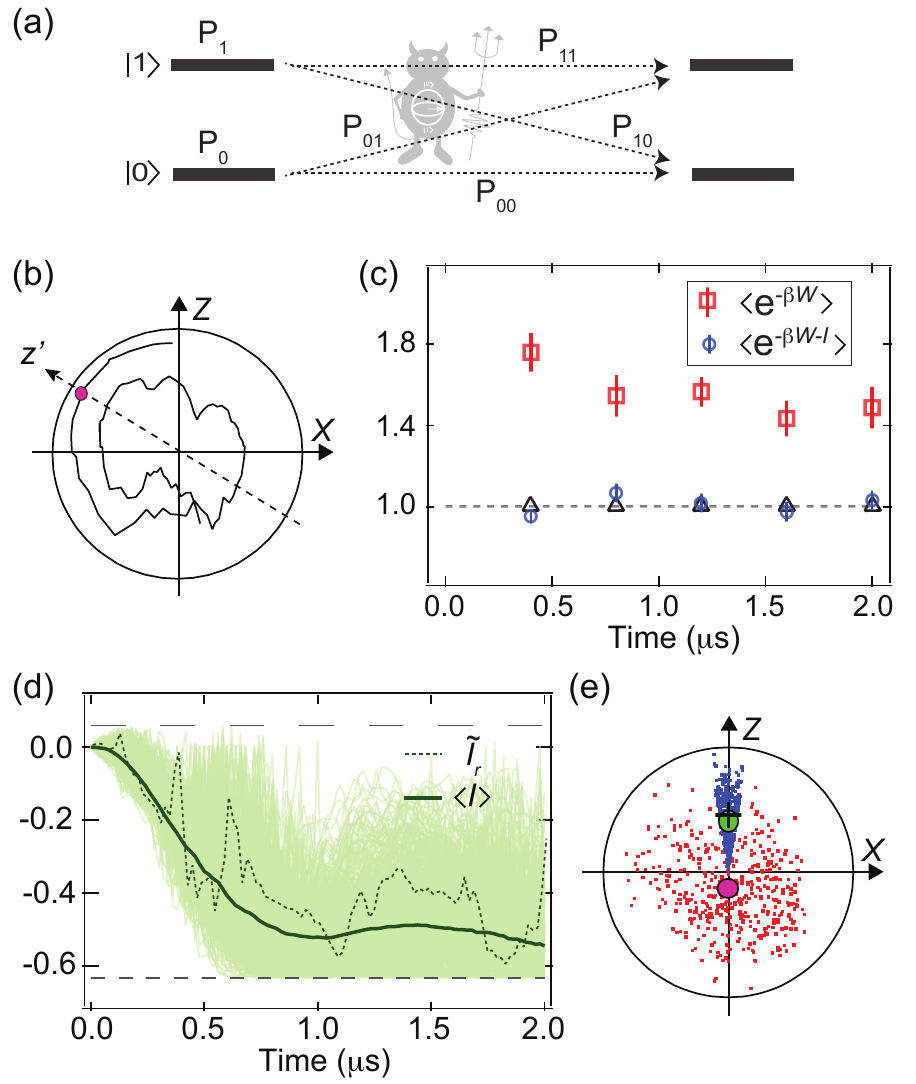}
\caption{ Experimental test of the quantum fluctuation theorem. (a) Transition probabilities are calculated using the two projective measurements in Steps 2 and 5. (b) By tracking a single quantum trajectory we calculate the information change  (\ref {Mut_inf2}). The diagonal basis $z'$ at time $t$ is indicated.   (c) Combining the transition probabilities and the information change we verify the quantum fluctuation theorem \eqref{jarcq} (round markers), however if the information is ignored ($I=0$) the fluctuation theorem is not valid (square markers).  The fluctuation theorem is also verified at the level of single quantum trajectories where the projective measurements are not used to determine transition probabilities (triangle markers).  (d) Information change, $I$, Eq.~\eqref{Mut_inf1}, along single quantum trajectories; the dotted curve shows a typical information change obtained by Eq.~\eqref{Mut_inf2}. The background color shows the distribution of information change for many trajectories. The solid curve is the average information change, $\langle I \rangle$, Eq.~\eqref{Mut_inf3}. Two dashed line shows the Shannon entropy of initial state $H(0)$ (coarse dash)  and $H(0)-\ln(2)$ (fine dash) which indicate the maximum and minimum limit for information change for a given initial state. (e) Red (Blue) dots show the qubit state distribution obtained by trajectories at $t=2\ \mu$s right before (after) the feedback rotation and the average of this distribution indicated by the red (green) circle. The cross indicates the reconstruction of the state of the qubit after the feedback using projective measurements. }
\label{fig3}
\end{figure}

In order to test the quantum fluctuation theorem for the considered two-level system, we write it explicitly as,
\begin{eqnarray}
\langle e^{-\beta W - I} \rangle 
&=& P_0(0)P_{00}(\tau)e^{-I_{00}} +  P_1(0)P_{11}(\tau)e^{-I_{11}}\nonumber \\
&+&  P_0(0) P_{10}(\tau) e^{+\beta-I_{10}}  + P_1(0) P_{01}(\tau)  e^{-\beta-I_{01}} \nonumber\\
&=& 1,
\label{jarcq}
\end{eqnarray}
with $\Delta F=0$ since the initial and final Hamiltonians are here the same. The initial occupation probabilities for ground and excited states are respectively given by $P_{0}(0)=1/(1+e^{-\beta})$ and $P_{1}(0)= e^{-\beta}/(1+e^{-\beta})$, corresponding to the initial thermal distribution. Note that we work in units where $\hbar \omega_\mathrm{q}=1$. We determine the transition probabilities $P_{ij}(\tau)$, $i,j = \{0,1\}$, from the results of the projective measurements performed in Steps 2 and 5, following the two-point measurement scheme \cite{talk07}, as illustrated in Fig.~\ref{fig3}a. We further evaluate the information term $I_{ij}= \ln P_{i}(\rho_{\tau|r}) - \ln P_j(\rho_0)$ from the recorded quantum trajectory according to Eq.~\eqref{Mut_inf1} (Fig.~\ref{fig3}b). In Fig.~\ref{fig3}c (round markers), we show the experimental result for Eq.~\eqref{jarcq}  for $\beta=4$ for five different protocol durations $\tau$. We observe that the generalized quantum Jarzynski equality with feedback is satisfied. However, the fluctuation theorem  is violated (square markers), as expected, when the information exchange is not taken into account.

Every measured trajectory contains a complete set of information by which the expectation value of any (relevant) operator can be calculated. In particular, the transition probabilities $P_{ij}(\tau)$ may be determined directly from the weak measurement data, instead of the outcomes of the two projective measurements. To establish the consistency between the two approaches, we rewrite the quantum fluctuation theorems with feedback \eqref{jarcq} as,  
\begin{eqnarray}
\langle e^{-\beta W-I} \rangle &=& P_0(0)P_{0}(\tau)e^{-I_{00}} +  P_1(0)P_{1}(\tau)e^{-I_{11}} \nonumber\\
&+ &P_0(0) P_{1}(\tau) e^{+\beta-I_{10}}  + P_1(0) P_{0}(\tau)  e^{-\beta-I_{01}} \nonumber \\
&=&1,
\label{jarq}
\end{eqnarray}
with the respective final ground and excited states populations $P_{0}(\tau)= P_0(\rho_{\tau|r})=(1+z(\tau))/2$ and $P_{1}(\tau)= P_1(\rho_{\tau|r})=(1-z(\tau))/2$, along single quantum trajectories. All these quantities  are obtained from the quantum state tracking in Step 3 and the consequent feedback rotation in Step 4. Figure~\ref{fig3}b (triangular markers) shows that Eq.~\eqref{jarq} is verified in our experiment and  that the two approaches are thus indeed consistent. 

Figure \ref{fig3}d shows the evolution of the information $\tilde I_r$  along single quantum trajectories calculated from Eq.~(\ref{Mut_inf2}). The probabilities in Eq.~\eqref{Mut_inf2} are evaluated at each time step in the diagonal basis $z'$, as illustrated in Fig.~\ref{fig3}b \cite{funo13}. Figure~\ref{fig3}e further exhibits the last point of 400 trajectories before (after) feedback rotation in red (blue). The red (green) circles indicate the expectation $\langle z \rangle$ before (after) the feedback rotation calculated using weak measurements. On the other hand, the black cross represents an independent evaluation of $\langle z \rangle$  from the second projective measurement data. The good agreement between the green circle and the black cross validates that feedback rotations are properly executed.

We next study the information dynamics of the mean information $\langle I\rangle$, Eq.~\eqref{Mut_inf3}, averaged over many trajectories, and the transition from information gain to information loss. In Fig.~\ref{fig3}d (solid curve), we observe that the mean information $\langle I\rangle$ for $\beta =4$ averaged over 400 trajectories is negative. This negativity of $\langle I\rangle$ is a consequence of both the quantumness of the dynamics, which generates states with coherent superpositions of the eigenstates of the measured observable,    and of the quantum backaction of the measurement \cite{jaco14}. For a classical measurement for which the density matrix commutes with the measurement operator, $\langle I\rangle$ reduces to the (positive) mutual information between the measurement result and the ensemble made up of the eigenstates of the density matrix. On the other hand, the quantum backaction of the inefficient measurement disturbs the state of the system and reduces our knowledge about it. When this information loss is larger than the information gained through the measurement, the total information exchanged is negative, as seen in the experiment.


\begin{figure}
	\centering
	\includegraphics[scale=0.9]{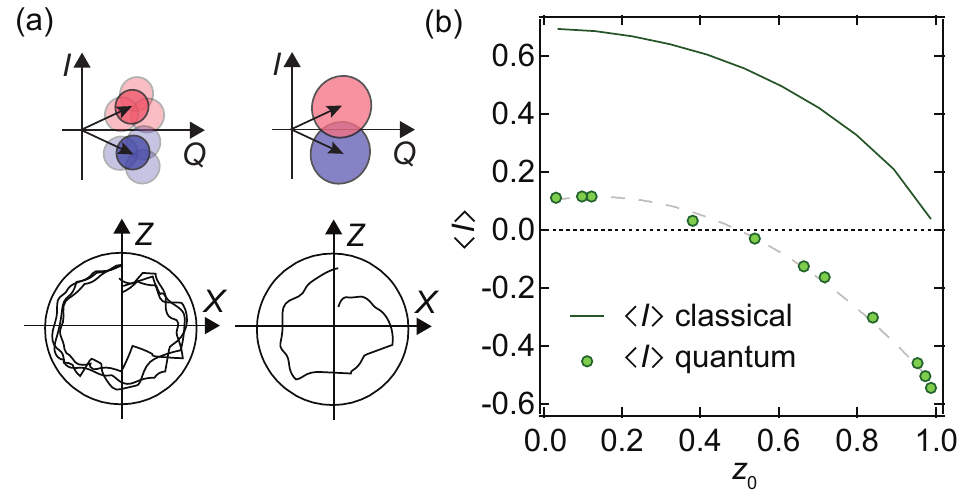}
	\caption{Transition from information gain to information loss.  (a) Classical and quantum noise corresponding to the unknown detector configuration randomly shifts the probe tone (shown here as a coherent state in the quadrature ($I$--$Q$) space of the electromagnetic field) and degrades the total efficiency of the measurement.  The inefficient unraveling of the SME is the statistical average of possible unravelings corresponding to different detector configurations. (b) Transition from information gain to information loss by changing the purity of the initial state of the system controlled by its temperature via $z_0= \tanh(\beta/2)$. The dashed line indicates a linear dependence of the information exchange on initial state purity and the solid line indicates the Shannon entropy of the initial state which is obtainable only with projective measurements corresponding to the Lanford-Robinson bound \cite{lanf68}.}
	\label{fig4}
\end{figure}

In our setup, the finite efficiency of the measurement, as shown in Fig.~\ref{fig4}a, follows from the fact that the detector signal is affected by classical and quantum noise which induces random shifts of the readout value. The resulting  uncertainty about the state of the detector determines the information loss \cite{funo13}. Meanwhile, the information gain may be controlled by the purity of the initial state, that is, by its temperature.   If we parametrize the initial thermal state as $\rho_0 = \left( {\bf 1} + z_0 \sigma_z \right)$/2, with $z_0 =\langle z \rangle|_{t=0} = \tanh{\beta/2}$, the information gain, and in turn the averaged information $\langle I \rangle$, is a monotonically decreasing function of $z_0$, as seen in Fig.~\ref{fig4}b. The transition to  $\langle I \rangle <0$ happens for sufficiently pure initial states, that is, for sufficiently large $z_0$, when the initial entropy of the system is low enough so that the $I_{\rm loss} $ induces by the measurement can overcome $I_{\rm gain}$. In the limit $\beta \rightarrow \infty$, the initial state would reduce to a pure state, corresponding to  $I_{\rm gain}=0$.

\textit{Conclusion}---We have experimentally realized a quantum Maxwell's demon using a continuously monitored driven superconducting qubit. By determining the information gathered by the  demon by tracking individual quantum trajectories of the qubit, we have first verified the validity of a quantum fluctuation theorem with feedback by using both a weak-measurement approach and the two-projective measurement scheme. In doing so, we have established the consistency of the two methods. We have further investigated the dynamics of the averaged  information exchanged with the demon and demonstrated that it may become negative, in stark contrast to the classical mutual information which is always a positive quantity. Because of the combined effect of the quantum coherent dynamics and of the quantum backaction of the imperfect measurement,  the description  of the demon  thus  requires quantum information.

\begin{acknowledgements} We acknowledge P.M. Harrington, J.T. Monroe, and D. Tan for discussions and sample fabrication. We acknowledge research support from the NSF (Grant  PHY-1607156) the ONR  (Grant  12114811), the John Templeton Foundation,  and the EPSRC (Grant EP/P030815/1). This research used facilities at the Institute of Materials Science and Engineering at Washington University. 
\end{acknowledgements}



\begin{thebibliography}{41}
\bibitem[{sam()}]{sampleparameters}
\bibinfo{note}{The experimental parameters are characterized by the qubit
  frequency $\omega_\mathrm{q}/2\pi = 5.11$ GHz, cavity frequency
  $\omega_\mathrm{c}/2\pi = 6.89$ GHz, Rabi frequency $\Omega_\mathrm{R}/2\pi =
  0.8$ MHz, dispersive coupling rate $\chi/2\pi = -0.3$ MHz, cavity bandwidth
  $\kappa/2\pi = 5.4$ MHz, measurment strength $k/2\pi = 51$ kHz, quantum
  efficiency $\eta=0.3$, and qubit coherence properties $T_1 = 5\ \mu$s, $T_2^*
  = 5\ \mu$s}.
\end{thebibliography}

\end{document}